\documentclass[11pt,draftcls,onecolumn]{article}
\usepackage{multicol,caption}
\usepackage{multirow}

\usepackage[a4paper, total={7in, 9in}]{geometry}
\usepackage{graphicx}
\usepackage{adjustbox}
\usepackage{changepage}
\usepackage{enumitem}
\usepackage{listings}
\usepackage{amsmath}
\usepackage{color}
\usepackage{authordate1-4}
\usepackage{rotating}
\usepackage{pdfpages}
\usepackage{amsfonts}
\usepackage{tikz}
\usepackage{bbm}
\tikzset{
	treenode/.style = {shape=rectangle, rounded corners,
		draw, align=center,
		top color=white, bottom color=blue!20},
	root/.style     = {treenode, font=\Large, bottom color=red!30},
	env/.style      = {treenode, font=\ttfamily\normalsize},
	dummy/.style    = {circle,draw}
}

\newenvironment{Figure}
{\par\medskip\noindent\minipage{\linewidth}}
{\endminipage\par\medskip}

\newcommand{\bc}{\begin{center}}
\newcommand{\ec}{\end{center}}
\newcommand{\tab}[1][1cm]{\hspace*{#1}}

\begin{document}
	\title{Patterns in demand side financial inclusion in India \\ \large{An inquiry using IHDS Panel Data}}
	\author{Vinay Reddy Venumuddala \\ PhD Student in Public Policy\\ Indian Institute of Management Bangalore}
	\date{}
	\maketitle

\begin{abstract}
In the following study, we inquire into the financial inclusion from a demand side perspective. Utilizing IHDS round-1 (2004-05) and round-2 (2011-12), starting from a broad picture of demand side access to finance at the country level, we venture into analysing the patterns at state level, and then lastly at district level. Particularly at district level, we focus on agriculture households in rural areas to identify if there is a shift in the demand side financial access towards non-agriculture households in certain parts of the country. In order to do this, we use District level ``Basic Statistical Returns of Scheduled Commercial Banks'' for the years 2004 and 2011, made available by RBI, to first construct supply side financial inclusion indices, and then infer about a relative shift in access to formal finance away from agriculture households, using a logistic regression framework. 
\end{abstract}

\section{Introduction}
Real improvement in income prospects from certain occupations, often is highly dependent on availability of relevant financial services within reach. While the importance of access to such services is felt almost uniformly across the country, nature of such an access is marked with stark disparities. The demand side pattern of such disparities, looking from the vantage point of a typical household, first emerges from identifying whether the access to financial services is from formal or informal sources. Financial products availed from informal sources are often considered to be costlier\footnote{More in terms of the rates of interest at which loans are offered}, and therefore more risky in comparison to those availed from formal sources. Second, from the level of a geographical region, this pattern may be expounded further by identifying differences in access to finance across rural or urban areas. Third, moving further away from the ground, a holistic but rather coarse pattern can be observable across districts within each state, or across states in the country.\\

 Existing literature on evaluating financial inclusion in India, are primarily driven by available data on supply side indicators. A seminal work on measuring financial inclusion at state level utilizing supply side information is proposed by \cite{chakravarty2013financial}. However, the efficacy of this supply side information used in \cite{chakravarty2013financial} is limited to capturing the demand side information rather coarsely\footnote{Although demand side information in terms of observing deposit/credit accounts per 1000 people are integrated into their \cite{chakravarty2013financial} index, identifying patterns in financial inclusion however is restricted only till the level of a state and not further below.}, at the level of a state. Nevertheless, building from this seminal work, we try to make a rather tiny effort, to understand financial inclusion from a demand side perspective. Further we try to extend our analysis (only to some extent)\footnote{We only remain with around 104 districts out of about 630+ districts, due to sample size limitations} upto the level of a district.\\
 
For the purpose of this study, we use the two rounds of IHDS (first in 2004 and second in 2011), to obtain information on household level access to formal/informal finance\footnote{We treat access to loans from `Banks and other government agencies' as Formal and `Other sources' as Informal. This information about the institution from which loan has been accessed from, is based on a household's largest loan taken over the past five years from the respective survey date(s) corresponding to each round}. Between the two rounds, we particularly focus our attention on the impact of supply side financial inclusion, in reaching to agriculture vis-a-vis non-agriculture households in rural areas across districts.

\section{Motivation}
In the Introductory chapter of the book, ``Talking financial inclusion in Liberalized India'', by Sriram \cite{sriram2017talking}, progress of financial inclusion in the country has been elucidated, under four major phases. With state led cooperatives, the first phase of financial inclusion is marked with a major focus on rural and agriculture households. In the next phase, the focus shifted to expanding the overall banking outreach in the country. Policy of social banking along with nationalization of banks, helped in providing banking to the hitherto un-banked regions. Literature also shows that this phase has been a major determinant of financial inclusion across states in the country \cite{chakravarty2013financial}, and also helped in addressing poverty to some extent \cite{burgess2005rural}. However, despite such expansion in banking outreach, regional disparity still remained in the country. South continued to dominate in terms of bank branches both in rural and urban areas, and other regions, particularly north-east, were lagging far behind. Third phase overlapping with the second, has given form to a new institutions called RRBs, with increased PSL targets for agriculture. Lastly, the fourth phase which is still ongoing, is marked by a gradual withdrawal of state, and promoting market forces to operate in the financial arena. However, state occasionally intervened through loan waivers, interest subventions, etc.., between banks and its customers, particularly for the rural poor and agriculture households. According to \cite{chakravarty2013financial}, withdrawal of social banking, had an adverse impact on overall financial inclusion subsequent to that period. \\

One particularly important feature that essentially determined the nature of financial inclusion in the country, is its focus on expanding access to credit for agriculture households across country. Despite policies stipulating priority sector lending, branch licensing to cater to rural areas, have in letter focused on attending to such an expansion, the actual penetration is rather grim.  That the financial institutions, particularly banks, are slowly becoming harder to access for households dependent on agriculture, is slowly becoming an unfortunate but stylized fact across India. Despite growth in share of formal credit as a part of agriculture GDP, the overall supply of formal credit to agriculture as a percentage of total disbursal is going down (\cite{mohan2006agricultural},\cite{sriram2007productivity}). Even the share of short term credit of agriculture households, as a proportion of their total output value, is hardly close to 5\%, indicating that formal credit may have less than marginal effect on their occupational sustenance, despite the amount of risks that are involved in carrying out agriculture \cite{sriram2007productivity}. Research indicates that banking system needs to bridge the gap in provision of financial services between rural and urban areas, and more so in facilitating agriculture households, particularly the small and marginal farmers \cite{mohan2006agricultural}. There is a substantive necessity to devise suitable financial products for agriculture, particularly taking into account the factors from demand side \cite{sriram2007productivity}. \\

The following study is motivated by an understanding of the evolution of financial inclusion policies\footnote{Many of them intending to facilitate improved access to rural and particularly agriculture households at-least in letter if not spirit.}, and the importance attributed to reach out to particularly the poor agriculture households. However, we restrict the scope of our study in observing the current patterns in financial inclusion which overlaps more or less only with the fourth phase as indicated in \cite{sriram2017talking}. Utilizing data on Debt and Investment, from the two rounds of IHDS (first in 2004, second in 2011), we inquire into financial inclusion through a demand side perspective by broadly looking at the following questions at different geographical levels. 

\begin{itemize}
	\item \textbf{Q1}: At the country level, between 2004 and 2011, what is the pattern in access to formal finance observed across households with varying income/asset levels and depending on different occupations? Is there any observable shift in this pattern between the two rounds?
	\item \textbf{Q2}: How does such a pattern vary across states, and between the two rounds (Owing to insufficient sample size, here we employ occupational categorization of  households as principally depending on agriculture and non-agriculture activities)? 
	\item \textbf{Q3}: Is there any identifiable shift in access to formal finance more towards non-agriculture households, in rural areas, within different districts in the country? If there is such a shift, how does it manifest within agricutlure households, analysed from their income/asset levels?
\end{itemize}

\section{Data}
For our analysis at the country and state level, we construct IHDS Panel Data by merging both round 1 and 2. However, for the analysis at district level, we had to use the following secondary data sources together for constructing a panel with district as unit of analysis, in order to inquire into our question(s).
\begin{itemize}
	\item \textbf{IHDS-I (2004-05) and IHDS-II (2011-12):} We use IHDS (Indian Human Development Survey) household panel data to construct a measure indicating the relative gap in difference of proportion of agriculture and non-agriculture households in accessing finance from banks, over the two rounds.
	\item \textbf{Basic Statistical Returns of Scheduled Commercial Banks}: In order to arrive at the indicators for geographic and demographic penetration of scheduled commercial banks across districts, we use this data from RBI. We use data for years 2004, and 2011 to integrate it into the panel.
	\item \textbf{Census:} District level population, and area for measuring the banking indicators mentioned above, are taken from District level data of Census 2001 and 2011.
\end{itemize}

\section{Observations and Discussion}
\subsection{Pattern of financial access across occupations in the country}
%

In order to observe the pattern in access to finance at the country level, we use asset indicators and income ranks of households across occupations\footnote{All along the study, we only focus our analysis on households accessing finance either from formal/informal sources.} as the variables of interest. Asset Indicators are constructed in IHDS dataset, which essentially indicate the level of assets ranging between 0-31. In order to obtain a comparable income ranks across occupations, we construct a metric of standardized income rank within each occupation. In order to do this, we first rank the households within an occupation based on their income, then we standardize this rank between 0-100, essentially giving an ordered rank of households for each occupation as a percentile. While asset score reflects an overall position of the households across all occupations, percentile income ranks denote the relative ranking of households within each occupation. Figure-\ref{fig:Q1} shown below gives the box-plots of these percentile income ranks, and of assets across different occupations.

 
\begin{Figure}
	\captionsetup{font=scriptsize}
	\begin{center}
		\includegraphics[width=7in]{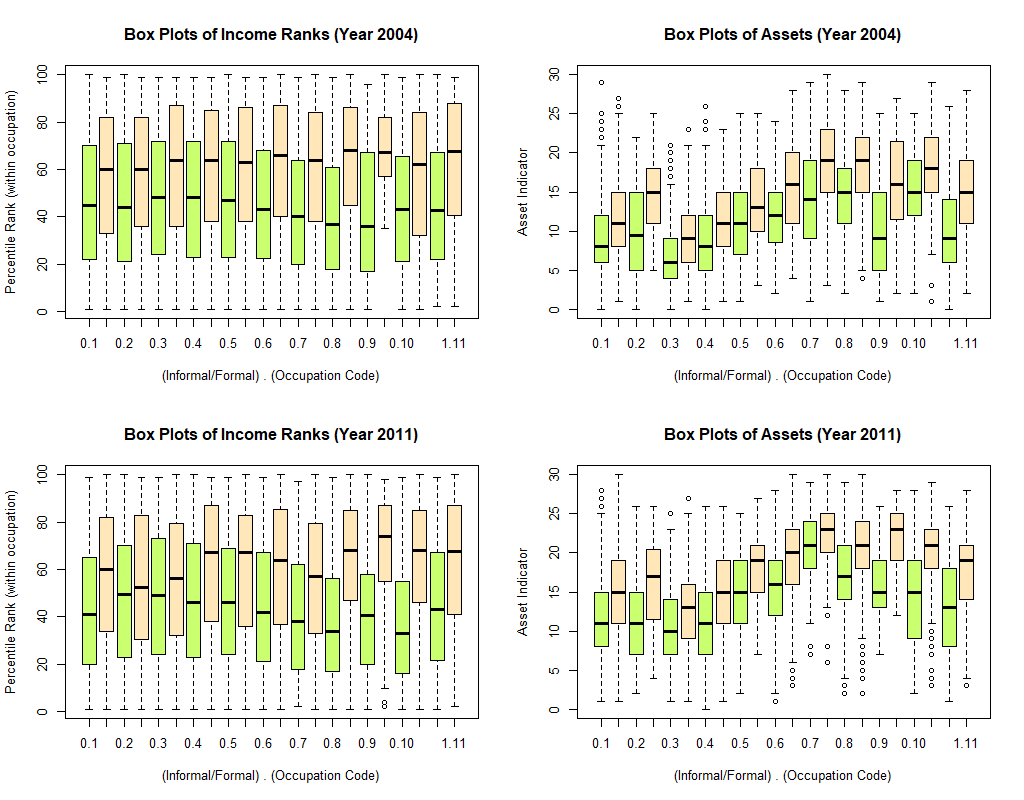}
		\captionof{figure}{Pattern of financial access across occupations.\\ \tiny{Occupational codes in order, ``(01) Cultivation'', ``(02) Allied agr'', ``(03) Ag labour'',``(04) Non-Ag labour'', ``(05) Artisan'', ``(06) Petty trade'',``(07) Business'',``(08) Salaried'',``(09) Profession'',``(10) Pension/rent'', ``(11) Others''}} 
		\label{fig:Q1}
	\end{center}	
\end{Figure}


It can be seen that households accessing formal finance have greater median value of asset indicators or income ranks, across all occupations. Observing the gap in median asset scores between formal and informal households across occupations, we see that the gap has increased for certain occupations (agriculture and allied activities, professional class and others), while it decreased in case of other occupations (like business). This indicates that the supply side financial inclusion, may have helped in bridging the gap between access to formal vs informal finance, in certain occupations more than others. Similar patterns can be found from the within occupation income ranks, where the gap in terms of percentile rank of households who are able to access formal vs informal finance, is widening in some occupations, while it has come down in case of others. Delving further into the factors that may have/haven't given occasion to this gap in access to finance over the years, may provide valuable policy inputs, especially in helping to devise suitable financial products across occupations. However, given the limited scope of our study (particularly with regard to the dataset used) we do not delve further into these aspects. 

\subsection{Pattern of financial access across states}
In order to observe the variation in financial access across states, we only use two major categories of occupations, agriculture and non-agriculture, owing to sample size \footnote{ See Figure-\ref{fig:Q2SS}, for the sample sizes in our dataset across states. This further reduced at district level as discussed in subsequent sections.}. For identifying the patterns across states, we use asset indicator alone as our variable of interest. Following figure-\ref{fig:Q2} gives the box plots of asset indicators for a set of X-variables in order \{\textit{0.0.State -$>$ Informal,Non-Agriculture; 1.0.State -$>$ Formal,Non-Agriculture; 0.1.State -$>$ Informal,Agriculture;  1.1.State -$>$ Formal,Agriculture}\}\footnote{Informal indicates those households whose major source of financial access is from informal sources, Similarly, formal indicates household's access to formal sources.}, for each state.\\

 From the box plots it can be seen that the gap in median asset score between households accessing formal vs informal finances, is widely varying across states, and even across households (on the basis of whether or not they are agricultural), within and across rounds. This varying pattern across states, and also changes over the years, calls for further investigating as to what may be the policies across states that may be giving occasion to these patterns, and the resulting changes over the years. \\
 
 In this study, however, we restrict ourselves on a rather conspicuous inference, that there is a visible variation in patterns observed across states, not only in terms of households getting access to formal finance or not; But also, about the status of households who are being able to access formal finance. Analysing the disparities between households who are able to access formal finance and those who are not, further along socio-economic, and particularly occupational heterogeneities, may give valuable inferences for policy making. Nevertheless, we are limited by the scope of this study in venturing into such an analysis.

\begin{Figure}
	\captionsetup{font=scriptsize}
	\begin{center}
		\includegraphics[width=7in]{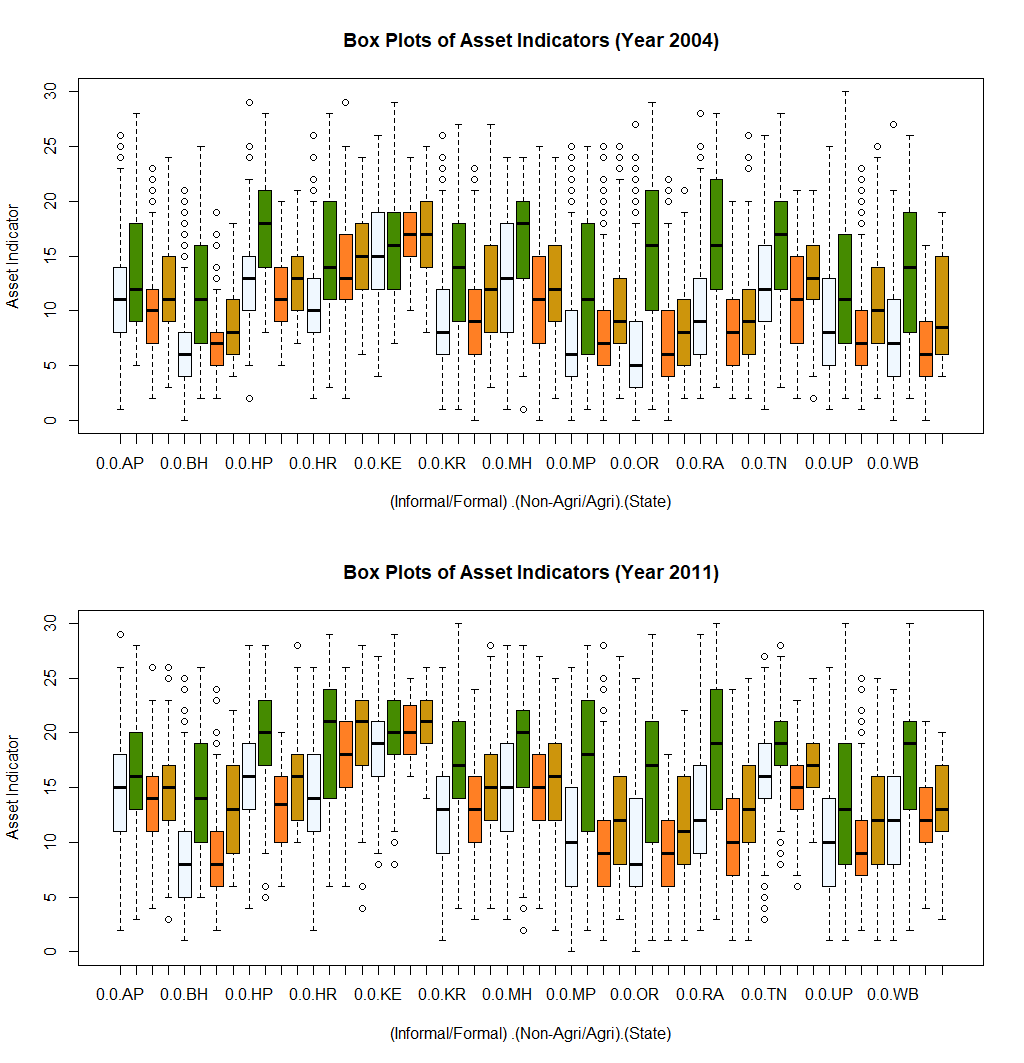}
		\captionof{figure}{Pattern across states} 
		\label{fig:Q2}
	\end{center}	
\end{Figure}

\begin{Figure}
	\captionsetup{font=scriptsize}
	\begin{center}
		\includegraphics[width=4.5in]{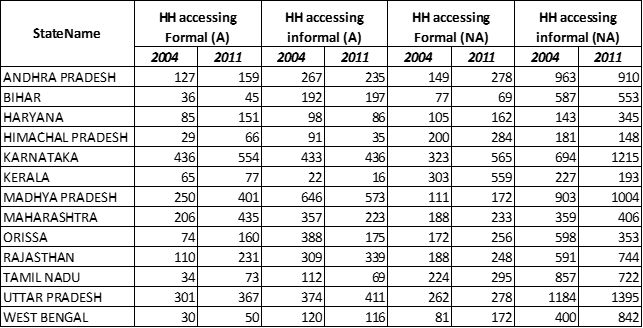}
		\captionof{figure}{Sample Sizes for each state in our dataset} 
		\label{fig:Q2SS}
	\end{center}	
\end{Figure}


\subsection{District Level Analysis - Inquiry into a possible shift in access to finance}
\subsubsection{Methodology and Results}
For each district, we consider the following indicator expressed in terms of difference between relative improvement in access to finance of agriculture vis-a-vis non-agricultural households, as our dependent variable. 
\begin{align*}
&\text{For each district $i$ we have the indicator as follows,}\\
&D_i = \mathbbm{1}\{p_{n\tau} - p_{n0} > p_{a\tau} - p_{a0}\},\\
&where,\\
&p_{nt} = \frac{\text{No.of Non-Agri Dependent Households Accessing Finance from banks}}{\text{Total Non-Agri Dependent Households Accessing Formal/Informal Finance}} \\
&p_{at} = \frac{\text{No.of Agri Dependent Households Accessing Finance from banks}}{\text{Total Agri Dependent Households Accessing Formal/Informal Finance}} \\
&\text{Above proportions are defined at t=$\tau$ (2011), and t=0 (2004)}
\end{align*}
The indicator function, essentially takes a value one when the relative increase in proportion of agriculture households accessing finance from banks increases less than that for non-agriculture households.\\

In order to differentiate districts along certain financial inclusion scale, we construct a rural financial inclusion index, using the approach proposed in \cite{chakravarty2013financial}. However, we only consider three\footnote{We consider Standardized indicators of Geographic penetration, Demographic penetration and number of deposit accounts per 1000 population, and then take the average to compute our financial inclusion index for rural areas (Approach is based on the Financial Index developed by \cite{chakravarty2013financial}, with r-value = 1, and number of indicators are 3 instead of 6)} out of six indicators while framing our index, to reflect only a supply side measure of financial inclusion. 
\begin{align*}
&(RFI\_index) \ G_{it} = \text{Rural financial inclusion index}\\
&(relchRFI) \ RG_i = \frac{\Delta G_i}{G_{i0}} = \frac{G_{i\tau} - G_{i0}}{G_{i0}}
\end{align*}

In order to model the nature of relative shift in access to formal finance towards/away from agriculture households, we run four regressions (shown below). Our dependent variable is the probability that our indicator takes the value 1, or to put in words, ``Non-agriculture households are relatively better off in access to formal finance in comparison to agriculture households in rural areas''. Independent variables of interest are the initial and average levels of rural supply side financial inclusion indicators, and initial and average levels of inequality (measured based on asset score/ total income) within agriculture households. We use inequality within agriculture hosueholds of a district as independent variable mainly to observe if a higher inequality within agriculture households may be responsible for a shift in access to finance towards or away from agriculture households. If there is any such noticeable shift in favour of agriculture households, along with higher inequality, it may imply that the fruits of supply-side financial inclusion may have increasingly benefited the high-income/asset agriculture households. However, we couldn't find any statistically significant results, validating either for or against such a hypothesis.\\

We run the following regressions in order to find the contribution of rural supply side financial inclusion on the shift in access to finance (shift - as discussed above). 

\begin{align*}
&logit\{P(D_i = 1)\} = \beta_0 + \beta_1 RG_i  + \beta_2 IA1_{i0} + \beta_3 G_{i0} + \epsilon_i\\
&logit\{P(D_i = 1)\} = \beta_0 + \beta_1 RG_i + \beta_4 G_{iavg}  + \beta_5 IA1_{iavg} + \epsilon_i\\
&logit\{P(D_i = 1)\} = \beta_0 + \beta_1 RG_i  + \beta_2 IA2_{i0} + \beta_3 G_{i0} + \epsilon_i\\
&logit\{P(D_i = 1)\} = \beta_0 + \beta_1 RG_i + \beta_4 G_{iavg}  + \beta_5 IA2_{iavg} + \epsilon_i\\
\end{align*}
$IA1$ indicates asset based inequality within rural agriculture households of the district, and $IA2$ indicates income inequality within the same set of households for each district. Subscript $avg$ indicates the average of the respective covariate taken over the two time periods. Table-\ref{tab:reg} shows the results from our four logistic regression models. In the table, $Pindex$ indicates $P(D_i = 1)$. The dependent variable therefore is the probability that non-agriculture households are relatively better in terms of access to formal finance than agricultural households. Covariates $AgGini04$ and $avgAgGini$ are equivalent to $IA1_0 \ or \ IA2_0$ and $IA1_{avg} \ or \ IA2_{avg}$ respectively, depending on asset/income based inequality measured ( Table-\ref{tab:reg} indicates this along the column descriptions).

\begin{table}[!htbp] \centering 
	\caption{Logistic Regression Results} 
	\label{tab:reg} 
	\begin{tabular}{@{\extracolsep{5pt}}lcccc} 
		\\[-1.8ex]\hline 
		\hline \\[-1.8ex] 
		& \multicolumn{4}{c}{\textit{Dependent variable:}} \\ 
		\cline{2-5} 
		\\[-1.8ex] & \multicolumn{4}{c}{Pindex} \\ 
		& \multicolumn{2}{c}{Asset Based Inequality} & \multicolumn{2}{c}{Income Based Inequality} \\ 
		\\[-1.8ex] & (1) & (2) & (3) & (4)\\ 
		\hline \\[-1.8ex] 
		relchRFI & $-$1.187 & $-$1.578 & $-$1.317 & $-$1.817 \\ 
		& (1.033) & (1.170) & (1.024) & (1.156) \\ 
		AgGini04 & 2.127 &  & 0.425 &  \\ 
		& (3.540) &  & (1.468) &  \\ 
		RFI\_index04 & 4.544$^{*}$ &  & 4.593$^{*}$ &  \\ 
		& (2.446) &  & (2.509) &  \\ 
		avgRFI &  & 4.706$^{**}$ &  & 4.679$^{*}$ \\ 
		&  & (2.358) &  & (2.469) \\ 
		avgAgGini &  & 5.565 &  & $-$0.552 \\ 
		&  & (4.527) &  & (1.007) \\ 
		Constant & $-$0.203 & $-$0.852 & 0.027 & 0.479 \\ 
		& (0.931) & (1.000) & (0.949) & (0.772) \\ 
		\hline \\[-1.8ex] 
		Observations & 104 & 104 & 104 & 104 \\ 
		Log Likelihood & $-$54.818 & $-$53.877 & $-$54.957 & $-$54.507 \\ 
		Akaike Inf. Crit. & 117.635 & 115.755 & 117.913 & 117.015 \\ 
		\hline 
		\hline \\[-1.8ex] 
		\textit{Note:}  & \multicolumn{4}{r}{$^{*}$p$<$0.1; $^{**}$p$<$0.05; $^{***}$p$<$0.01} \\ 
	\end{tabular} 
\end{table}

However, these results need to be interpreted with great caution. In order to be able to understand the limitations better, we give a brief overview of how the data has been constructed for our study. 
\subsubsection{Panel Data Construction with district as unit}
\tab \textbf{Data-Set Creation}: Firstly, we merge census 2001 and 2011, on the basis of those districts which had the same area during both the rounds. We get around 649 districts with matched area across these two rounds. Then we separately merge banking statistics for the year 2004 (capturing information of close to 564 districts), and 2011 (capturing information of close to 632 districts), yielding data for 555 common districts\footnote{Please note that this attrition from 564 to 555 is purely due to mismatch while comparison of district names, we ignore this attrition, as we consider it to be purely random}. We now use OpenRefine software from Google, to create lookup tables for the district names matchable across Census, Banking Statistics, and IHDS. Having done this, we finally create a panel of district level banking statistics, area and population statistics, for close to 188 districts, whose names match with that of districts in IHDS.\\

\hspace{0.1cm} \textbf{Sample Reduction}: In order to be able to interpret the results from a set statistical reference, we chose those districts alone where the number of households accessing finance is atleast 20 in either of the rounds\footnote{Reducing further is drastically reducing our sample of districts particularly because of lower number of households accessing formal finance during round 1} (2004  or 2011). We therefore are left with only a total of about 104 districts, which match this constraint. Subsequently, we construct inequality among agriculture households within each district (from IHDS data), and the financial inclusion index for each district (from Banking Statistics Data), so as to adequately construct our independent variables in the regression. Dependent variable ($Pindex$) constructed using IHDS data, is also integrated into this dataset. Thereby we obtain a panel with data of 104 districts\footnote{List of these districts is provided in the Appendix} having all the relevant variables for our regression.

\subsubsection{Results Interpretation - District level}
Within its limitations, the above model roughly indicates that initial/average level of inequality computed either using assets/incomes of agriculture households, is not affecting the Probability that there is a greater shift towards agriculture/non-agriculture households. However, we can see that greater the supply side financial inclusion index, higher is this probability of shift. Which indicates a surprising result, that, it is in developed\footnote{In terms of higher supply side financial inclusion} regions where the proportion of non-agriculture households accessing formal finance 
is increasing in comparison to proportion of agriculture households (In both cases only among those households accessing either formal/informal finance). However, limited by sample size of our dataset at the district level, we may not be able to firmly conclude that this association is indeed significant. More robust checks, and particularly deploying bayesian hierarchical methods which account for neighbourhood effects, may help in throwing better light on the results. From these results, we only infer a vague hint that in rural areas of districts with better supply side financial inclusion, the relative shift of formal finance towards non-agriculture households is greater in comparison to agriculture households (on average). 

\section{Conclusion}
Above study is limited by the dataset used for analysis. Dataset with greater sample sizes particularly at district level; and capturing both access to formal finance and asset holdings/incomes may better reveal the patterns from a demand side view point. Nevertheless, in this study we made a small effort in observing recent patterns in access to finance from a demand side, using the IHDS rounds. We find that the difference in median asset scores of households accessing formal finance to that of informal ones, across all states in the country is always positive. However, this difference (median gap) is seen to vary with occupations, states, and across years. In order to investigate further at the district level, we construct a district level dependent variable ($Pindex$), which indicates whether or not the relative increase in access to formal finance (between the two IHDS rounds) had been better for non-agriculture households over agriculture households. Regressing on the levels of supply side financial inclusion, and the asset/income rank inequalities of agriculture households, we observe some vague hints of a pull back of formal financing institutions away from agriculture households. In districts with better supply side financial inclusion, we vaguely observe that, relatively, agriculture households are moving away from these institutions, in comparison to non-agriculture households. However, owing to data limitations, we may not be able to deny the possibility that these results may be just statistical artefacts. \\ 

From the policy perspective, it becomes extremely important to further investigate the factors underlying these movements either towards or away from formal financial institutions. Given there is a substantive need for devising suitable financial products particular for agriculture households, disentangling these factors becomes even more essential. Another important requirement is from the point of view of data. From the supply side, there is good amount of data provided by RBI on Banking penetration till the level of a district. However, from the demand side, there is little available data till the level of district. Analysing the movements towards or away from formal financial institutions, of households belonging to various occupations, with time, can help greatly in assessing the true manifestation of financial inclusion. \\

To conclude, although not sufficiently robust, the methodology adopted in the paper, can be helpful in providing a framework to look at the patterns of financial inclusion from demand side. 

\section{Appendix}
\begin{Figure}
	\captionsetup{font=scriptsize}
	\begin{center}
		\includegraphics[width=6in]{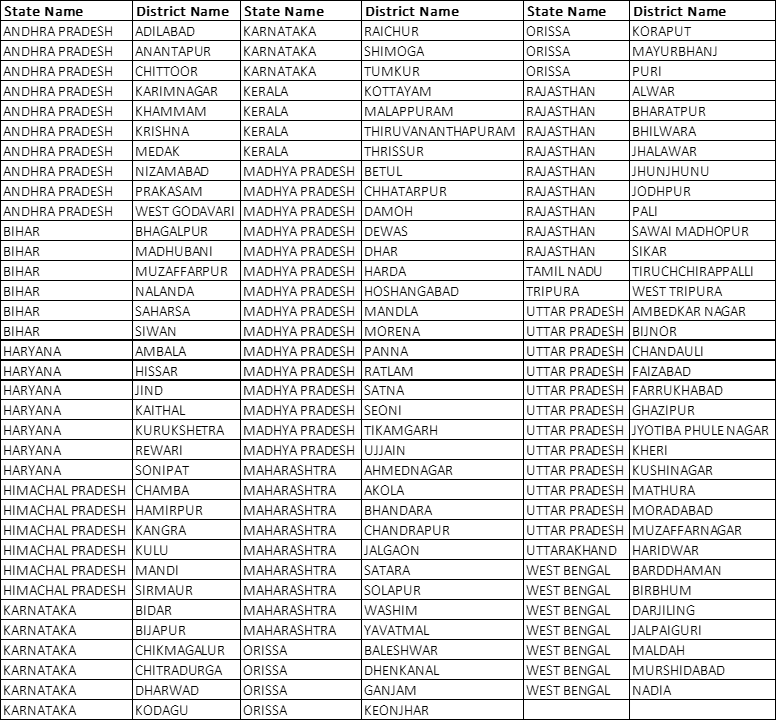}
		\captionof{figure}{List of District on which district level panel is constructed} 
		\label{fig:Dist}
	\end{center}	
\end{Figure}

\bibliographystyle{authordate1}
\bibliography{reflibrary}

\end{document}